\documentclass[twocolumn]{aastex63}

\usepackage{xcolor}

\newcommand{\magic}{MAGIC}
\newcommand{\lat}{\textit{Fermi}/LAT}
\newcommand{\gray}{$\gamma$-ray}
\newcommand{\gspaceray}{$\gamma$ ray}
\newcommand\txs{TXS~0506+056}

\received{\today}
\revised{\today}
\accepted{\today}

\shorttitle{Multi-wavelength campaign on \txs\ during 2017-2019}
\shortauthors{MAGIC Collaboration et al.}

\begin{document}

\title{Investigating the blazar \txs\ through sharp multi-wavelength eyes during 2017-2019 }

\author[0000-0001-8307-2007]{V.~A.~Acciari}
\affiliation{Instituto de Astrof\'isica de Canarias and Dpto. de  Astrof\'isica, Universidad de La Laguna, E-38200, La Laguna, Tenerife, Spain}
\author{T.~Aniello}
\affiliation{National Institute for Astrophysics (INAF), I-00136 Rome, Italy}
\author[0000-0002-5613-7693]{S.~Ansoldi}
\affiliation{Universit\`a di Udine and INFN Trieste, I-33100 Udine, Italy}\affiliation{also at International Center for Relativistic Astrophysics (ICRA), Rome, Italy}
\author[0000-0002-5037-9034]{L.~A.~Antonelli}
\affiliation{National Institute for Astrophysics (INAF), I-00136 Rome, Italy}
\author[0000-0001-9076-9582]{A.~Arbet Engels}
\affiliation{Max-Planck-Institut f\"ur Physik, D-80805 M\"unchen, Germany}
\author[0000-0002-4899-8127]{M.~Artero}
\affiliation{Institut de F\'isica d'Altes Energies (IFAE), The Barcelona Institute of Science and Technology (BIST), E-08193 Bellaterra (Barcelona), Spain}
\author[0000-0001-9064-160X]{K.~Asano}
\affiliation{Japanese MAGIC Group: Institute for Cosmic Ray Research (ICRR), The University of Tokyo, Kashiwa, 277-8582 Chiba, Japan}
\author[0000-0002-2311-4460]{D.~Baack}
\affiliation{Technische Universit\"at Dortmund, D-44221 Dortmund, Germany}
\author[0000-0002-1444-5604]{A.~Babi\'c}
\affiliation{Croatian MAGIC Group: University of Zagreb, Faculty of Electrical Engineering and Computing (FER), 10000 Zagreb, Croatia}
\author[0000-0002-1757-5826]{A.~Baquero}
\affiliation{IPARCOS Institute and EMFTEL Department, Universidad Complutense de Madrid, E-28040 Madrid, Spain}
% ...
 \collaboration{199}{(MAGIC collaboration)}
 \author{M. Hodges}
\affiliation{Owens Valley Radio Observatory, California Institute of Technology, Pasadena, CA 91125, USA}
 \author[0000-0002-2024-8199]{T. Hovatta}
\affiliation{Finnish Center for Astronomy with ESO (FINCA), University of Turku, FI-20014, Turku, Finland}
\affiliation{Aalto University Mets\"ahovi Radio Observatory, Mets\"ahovintie 114, 02540 Kylm\"al\"a, Finland}
 \author[0000-0001-6314-9177]{S. Kiehlmann}
\affiliation{Institute of Astrophysics, Foundation for Research and Technology-Hellas, GR-71110 Heraklion, Greece}
\affiliation{Department of Physics, Univ. of Crete, GR-70013 Heraklion, Greece}
\author[0000-0001-9200-4006]{I. Liodakis} \affiliation{Finnish Center for Astronomy with ESO (FINCA), University of Turku, FI-20014, Turku, Finland}
 \author[0000-0002-5491-5244]{W. Max-Moerbeck}
\affiliation{Departamento de Astronomía, Universidad de Chile, Camino El Observatorio 1515, Las Condes, Santiago, Chile}
 \author[0000-0001-5213-6231]{T. J. Pearson}
\affiliation{Owens Valley Radio Observatory, California Institute of Technology, Pasadena, CA 91125, USA}
 \author{A. C. S. Readhead}
\affiliation{Owens Valley Radio Observatory, California Institute of Technology, Pasadena, CA 91125, USA}
 \author[0000-0001-5704-271X]{R. A. Reeves}
\affiliation{Departamento de Astronomía, Universidad de Conceptión, Concepción, Chile}
 \collaboration{8}{(OVRO collaboration)}
\author[0000-0002-0393-0647]{A. Lähteenmäki}
\affiliation{Aalto University Mets\"ahovi Radio Observatory, Mets\"ahovintie 114, 02540 Kylm\"al\"a, Finland}
\affiliation{Aalto University Department of Electronics and Nanoengineering, P.O. BOX 15500, FI-00076 AALTO, Finland}
 \author[0000-0003-1249-6026]{M. Tornikoski}
\affiliation{Aalto University Mets\"ahovi Radio Observatory, Mets\"ahovintie 114, 02540 Kylm\"al\"a, Finland}
 \author[0000-0002-9164-2695]{J. Tammi}
\affiliation{Aalto University Mets\"ahovi Radio Observatory, Mets\"ahovintie 114, 02540 Kylm\"al\"a, Finland}
\collaboration{3}{(Mets\"ahovi collaboration)}
\author[0000-0001-7618-7527]{F. D'Ammando$^\star$}
\affiliation{INAF-Istituto di Radioastronomia, Via Gobetti 101, I-40129 Bologna, Italy}
\author[0000-0003-3779-6762]{A. Marchini}
\affiliation{Universit\`a di Siena, I-53100 Siena, Italy}
\correspondingauthor{M. Cerruti, E. Prandini, C. Righi,\\ K. Satalecka, N. Sahakyan, F. D'Ammando}
\email{contact.magic@mpp.mpg.de}

\begin{abstract}

The blazar \object{TXS~0506+056} got into the spotlight of the astrophysical community in September 2017, when a high-energy neutrino detected by IceCube (IceCube-170922A) was associated at the 3\,$\sigma$ level to a \gray\ flare from this source. This multi-messenger photon-neutrino association remains, as per today, the most significant one ever observed.  
\txs\ was a poorly studied object before  the IceCube-170922A event. To better characterize its broad-band emission, we organized a multi-wavelength campaign lasting 16 months (November 2017 to February 2019), covering the radio-band (Mets\"ahovi, OVRO), the optical/UV (ASAS-SN, KVA, REM, {\em Swift}/UVOT), the X-rays ({\em Swift}/XRT, {\em NuSTAR}), the high-energy \gspaceray s ($Fermi$/LAT) and the very-high-energy (VHE) \gspaceray s (MAGIC). In \gspaceray s, the behaviour of the source was significantly different from the 2017 one: MAGIC observations show the presence of flaring activity during December 2018, while the source only shows an excess at the 4\,$\sigma$ level during the rest of the campaign (74 hours of accumulated exposure); $Fermi$/LAT observations show several short (days-to-week timescale) flares, different from the long-term brightening of 2017. No significant flares are detected at lower energies. The radio light curve shows an increasing flux trend, not seen in other wavelengths. We model the multi-wavelength spectral energy distributions in a lepto-hadronic scenario, in which the hadronic emission emerges as Bethe-Heitler and pion-decay cascade in the X-rays and VHE \gspaceray s. According to the model presented here, the December 2018 \gray\ flare was connected to a neutrino emission that was too brief and not bright enough to be detected by current neutrino instruments. \\
\end{abstract}


\section{Introduction} \label{sec:intro}
On September 22, 2017, the IceCube neutrino observatory \citep{IceCube06} detected a 290 TeV neutrino (IceCube-170922A) from a direction consistent with the blazar \txs, which was found to be flaring in \gspaceray s with both \lat\ and \magic. \lat\ detected the blazar in a long-lasting (about six months long) flaring state, while \magic\ detected a fast flare about 10 days after the IceCube neutrino. The chance probability that the IceCube event and the flaring in the \lat\ (\magic) energy band happened simultaneously in space and time was estimated at the 3\,$\sigma$ (3.5\,$\sigma$) level \citep{0506science1}, and is, as per today, the strongest evidence for joint photon and neutrino emission from an active galactic nucleus (AGN). This multi-messenger photon-neutrino emission is the \textit{smoking gun} attesting for the presence of highly-relativistic hadrons in AGN jets. Its firm detection would unequivocally identify AGNs as (ultra-) high-energy cosmic-ray accelerators. Motivated by the detection of IceCube-170922A, the IceCube collaboration searched for additional neutrinos coming from \txs\ and identified a cluster of events in 2014-2015, significant at the 3.5\,$\sigma$ level \citep{0506science2}. Interestingly, this six-months-long neutrino flare was not accompanied by an enhanced electromagnetic activity, although the multi-wavelength coverage was limited due to the absence of any specific alert at that time, and relies only on information
from survey instruments. \\

\begin{figure*}[ht!]
\centering
\includegraphics[width=\textwidth]{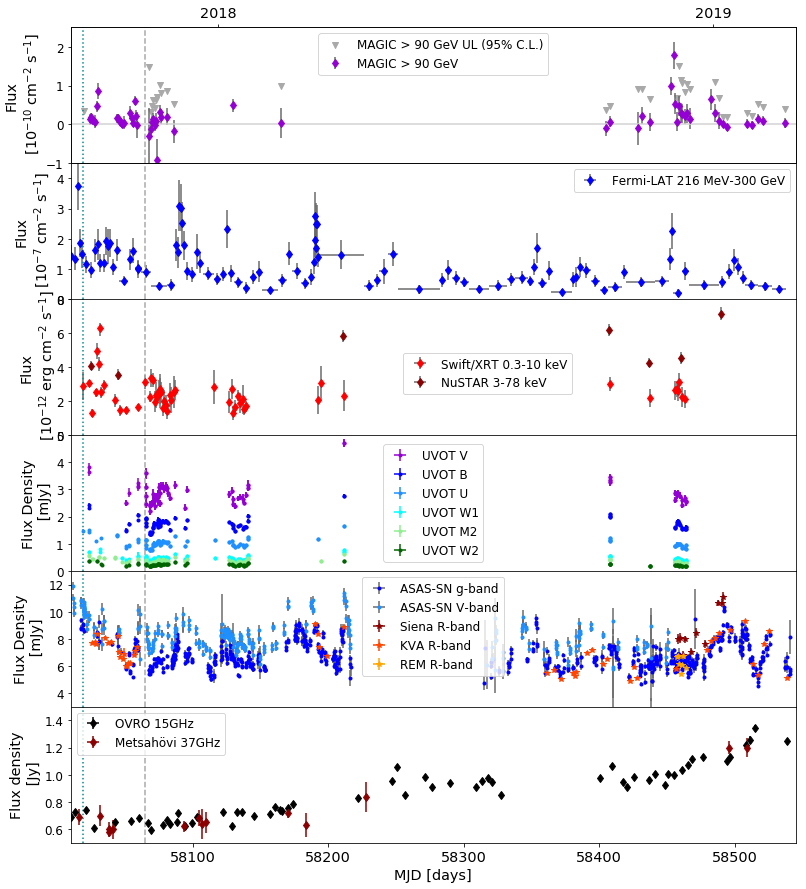}
\vspace{0.8cm}
 \caption{Multi-wavelength light curve of \txs\ during the 2017-2019 campaign. From top to bottom: MAGIC; $Fermi$/LAT; X-rays ({\em Swift}/XRT and {\em NuSTAR}); UVOT; optical (ASAS-SN, KVA, REM, Siena); radio (OVRO and Mets\"ahovi). The light-blue dotted line shows the arrival time of the IceCube-170922A neutrino, the grey dashed line marks the end of the data-set previously published in \citep{0506science1,MAGIC_TXS_paper2,2018MNRAS.480..192P}. All optical and X-ray fluxes are corrected from absorption by neutral material. MAGIC data are shown as observed, and are not corrected for pair-production absorption on the extragalactic background light. \\}
 \label{fig:mwl_lc} 
\end{figure*}

The two events, offering the first chance to test hadronic radiative models on multi-messenger data, have been extensively studied and modeled by several research groups. Standard blazar lepto-hadronic models, in which the neutrino emission is associated with the decay of pions produced in proton-photon interactions in the jet, are able to fit the electromagnetic spectral energy distribution (SED) and at the same time reproduce a neutrino rate consistent with the single event seen by IceCube in 2017. While single-zone models face the disadvantage of requiring a high proton power (often super-Eddington) to fit the data \citep{Cerruti19, Gao19}, solutions that make use of external target photon fields seem more promising \citep{MAGIC_TXS_paper2, Keivani18, Petropoulou20}. Another important conclusion from the IceCube-170922A modeling efforts is that pure hadronic models, in which the \gray\ emission is dominated by proton-synchrotron radiation, are excluded. \\

On the other hand, the 2014-2015 neutrino-only flare challenges current blazar emission models. Pion decay injects neutrinos together with photons and electrons/positrons in the emitting region. The latter radiate synchrotron and inverse-Compton photons (via a e$^\pm$-pair cascade that transfers the power to lower frequencies). This photon emission associated with the neutrino one has to be significantly suppressed in order to reproduce the 2014/2015 event. Alternatively, that photon emission can be channeled into the MeV band, where the upper limits are less constraining \citep[see][]{Rodrigues19}. This might suggest the existence of a dark neutrino emitting region, opaque to photons, and physically separated from the blazar emitting region that dominates the electromagnetic SED \citep{Reimer19, Xue19}. 

A possible solution could also be the \textit{"cosmic-ray-neutral-beam"} scenario, in which neutrons produced in proton-photon interactions escape the region and travel further down in the jet before interacting again and producing neutrinos, while the associated cascade emission is isotropized in the larger-scale jet \citep[see][]{Murase18, Zhang20}. Another alternative is represented by interactions of the relativistic jet with obstacles, such as clouds or stars: in this case neutrinos can be efficiently produced in proton-proton interactions \citep[see][]{Sahakyan18, Banik20} without strong ambient photon fields triggering electron-positron cascades. \\

Although the first candidate neutrino source is a blazar, it is important to underline that it is unlikely that \gray\ blazars represent the bulk of the diffuse neutrino background \citep{Aartsen172LAC, Hooper19}. With 10 years of IceCube data, there are still no sources individually significant at more than 5\,$\sigma$ in the neutrino sky \citep{Aartsen2010years}. The hottest spot in the all-sky scan (5\,$\sigma$ pre-trials, reduced to  $1.3 \sigma$ post-trials) is consistent with the AGN NGC~1068 (Seyfert galaxy with starburst activity); in the source catalog search, the most significant one is again NGC~1068 (4.1\,$\sigma$ pre-trial; 2.9\,$\sigma$ post-trials), while \txs\ is significant at the 3.6\,$\sigma$ level (pre-trial).
ANTARES, a neutrino telescope located in the northern hemisphere \citep{Antares}, reports 3FGL~J2255.1+2411 as the most significant blazar among the $Fermi$/LAT sample  \citep{AntaresPS}, significant at the 2.3\,$\sigma$ level (combined space and time probability) and at the 2.6\,$\sigma$ level when considering information from IceCube, that also detected a high-energy neutrino consistent with this source (IceCube-100608A). \\

\begin{figure*}[ht!]
    \centering
    \includegraphics[width=\columnwidth]{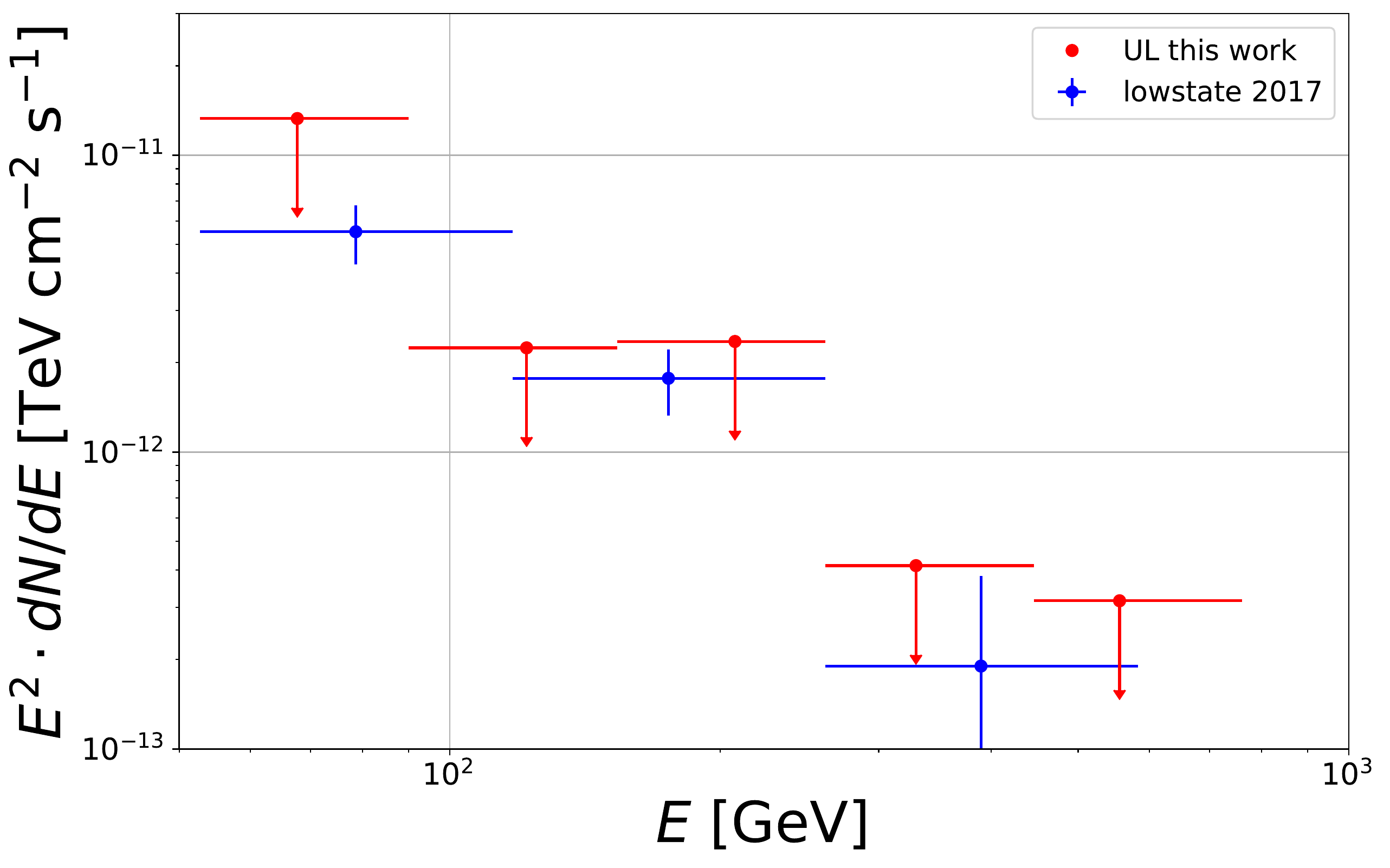}
    \includegraphics[width=\columnwidth]{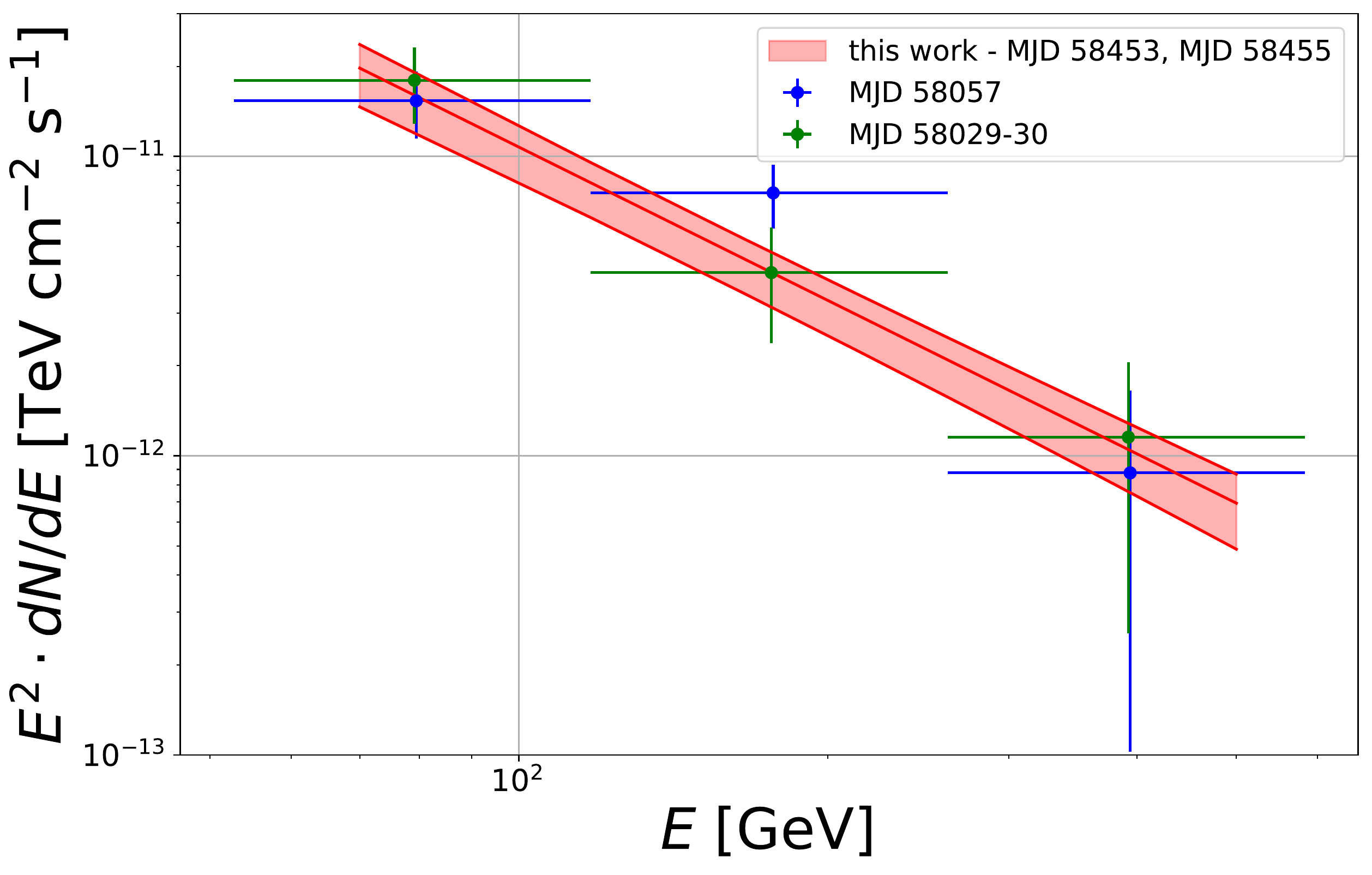}
    \caption {MAGIC spectral energy distribution as measured during this campaign (red). Low state is shown on the $left$ and the flare on the $right$. Overlaid are the points from the low state ($left$, blue) and flares ($right$, blue and green) observed in 2017 \citep{MAGIC_TXS_paper2}.\\}\label{fig:magic_sed}
\end{figure*}

\txs\ was a very poorly studied object before 2017. The interest triggered by its association with a high-energy neutrino prompted a measurement of its redshift \citep[z=0.337, ][]{Paiano18}, as well as several multi-wavelength follow-ups. It is remarkable that the source stands out among the blazar population as an atypical high-luminosity/high-synchrotron-peak blazar \citep{Padovani19}, with a synchrotron peak frequency located at about $10^{15}$ Hz \citep[right at the divide between intermediate and high-frequency-peaked blazars, see ][]{Abdo2010} and a peak luminosity of about $10^{47}$ erg s$^{-1}$ (a value more typical of flat-spectrum radio quasars). High-resolution VLBA radio images show possible signs of deceleration of the jet and/or a spine-layer structure within 1 mas from the mm-VLBI core of the source, as well as an apparent limb brightening \citep{Ros20}. After the 2017 multi-messenger event, follow-up observations have been conducted in the TeV band \citep[with the MAGIC and VERITAS telescopes, see][]{MAGIC_TXS_paper2, Veritas0506}, and in the optical-infrared band \citep{Hwang20}. \\

The absence of archival observations of \txs\ (besides survey instruments such as \lat\ and OVRO) means that, in 2017, it was not possible to estimate duty cycles of the source at various wavelengths, or estimate, for example, how exceptional was the \gray\ flare detected by MAGIC, or if the X-ray flux during the neutrino event was also unusually high.  To significantly improve our knowledge of this unique source, we started a multi-wavelength campaign during the 2017-2019 observing season, including data from radio to very-high-energy (VHE, $E>100$ GeV) \gspaceray s. Results from this campaign are presented in this paper and organized as follows. In Section \ref{sec:obs} we present the data analysis, from high to low frequencies (MAGIC, $Fermi$/LAT, {\em NuSTAR}, {\em Swift}/XRT, {\em Swift}/UVOT, optical telescopes, Mets\"ahovi 14\,m, OVRO 40\,m); in Section \ref{sec:var} we discuss the observed multi-wavelength variability of the source; in Section \ref{sec:modeling} we present the multi-messenger modeling of the observations; discussion of the results and conclusions are in Section \ref{sec:concl}. \\

\begin{table*}[t!]
\caption{MAGIC measurements of \txs}
\label{tab_param}
\centering
\begin{tabular}{l|ccc}
\hline
\hline
\textbf{Period} & \textbf{Exposure} & \textbf{Significance} & \textbf{Flux${>90\ \textrm{GeV}}$} \\
 & \footnotesize{[h]}&  &  \footnotesize{[10$^{-11}$ cm$^{-2}$ s$^{-1}$]} \\
 \hline
 MJD 58453 & 2.5 & 3.8\,$\sigma$ & 9.8$\pm$2.5  \\
 MJD 58455 & 1.8 & 5.4\,$\sigma$ & 18.0$\pm$3.4 \\
 MJD 58453-55 & 4.3 & 6.7\,$\sigma$ & 12.5 $\pm$ 1.8 \\
 Rest & 74.4 & 4.0\,$\sigma$ & 2.7$\pm$2.1  \\
 \hline
 \hline
 \end{tabular} \label{tab:magic_flux}
\end{table*}

\section{Observations and Data Analysis} \label{sec:obs}
This paper is based on an extended multi-wavelength data set collected from November 2017 till February 2019, i.e. giving continuation to the observations published in \cite{MAGIC_TXS_paper2}.  
Dedicated monitoring observations were performed with {\em Swift}/XRT \citep{SwiftXRT} and {\em Swift}/UVOT \citep{UVOT}, as well as {\em NuSTAR} \citep{harrison13}. They were coordinated to maximise the simultaneity of the observations with MAGIC.
MAGIC observations were usually accompanied by the KVA \citep{2018A&A...620A.185N} optical telescope, and additional optical measurements were performed with REM \citep{REM1,REM2}, and the University of Siena telescope. \txs\ is also systematically monitored by the ASAS-SN project \citep{2017PASP..129j4502K} and the radio telescopes OVRO \citep{2011ApJS..194...29R} and Mets\"ahovi.\\

Fig.\ref{fig:mwl_lc} shows the observed light curves from the VHE \gspaceray s to the radio band. To ease the comparison with the previously published results, we include in the same plot data from September and October 2017, i.e. the multi-wavelength campaign that followed the IceCube neutrino alert \citep{0506science1,MAGIC_TXS_paper2,2018MNRAS.480..192P}. 
The following sub-sections describe the data analysis for each of the instruments involved in the observing campaign.\\

\subsection{MAGIC}
\label{sec:magic}
The observations at VHE were performed with the Major Atmospheric Gamma-ray Imaging Cherenkov (MAGIC) stereoscopic system of telescopes, located at the Observatorio del Roque de los Muchachos, on the Canary island of La Palma, Spain. Each of the two telescopes has a 17m-diameter mirror and the system can achieve a sensitivity of 0.66\% of the Crab Nebula flux above 0.2 TeV in 50\,h of observations \citep{MAGICsens}.\\

Between November 2017 and February 2019 a total of $\sim$79 hours of good quality data, at zenith angles between 5$^\circ$ and 50$^\circ$, were collected. 
The observations were performed only in "dark" conditions (i.e. not affected by moonlight\footnote{Moonlight is defined as camera illumination resulting in the measured current higher than 2000 $\mu$A}) and in wobble mode \citep{Fomin94}. 
The data have been analysed using the MAGIC Analysis and Reconstruction Software \citep[MARS,][]{Moralejo09,Zanin13}.
When necessary, the flux values are corrected for the atmospheric extinction due to clouds and aerosols using the LIDAR system at the MAGIC site \citep{2014arXiv1403.3591F}. The resulting daily binned light curve of integral fluxes or flux upper limits above 90 GeV is presented in the top panel of Fig.\ref{fig:mwl_lc}. The upper limits are calculated for all observations with significance below 2\,$\sigma$, at the 95\% confidence level, using the method of \citet{2005NIMPA.551..493R}, and assuming a 30$\%$ systematic error on the signal efficiency, as nuisance parameter.\\ 

Table \ref{tab:magic_flux} shows the summary of the \txs\ flux levels as measured by MAGIC: two flaring episodes are highlighted: on December 1, and 3, 2018 (MJD 58453 and 58455); the remaining period is considered as \textit{low state}. The flux levels in table \ref{tab:magic_flux} are calculated assuming a power-law spectrum, with photon indices of 3.7 $\pm$ 0.2 for the flare and 3.5 $\pm$ 0.4 for the \textit{low state} period. The typical systematic uncertainties of MAGIC spectral measurements can be divided into: $<$ 15\% in energy scale, 11-18\% in flux normalization and $\pm 0.15$ for the energy spectrum power-law slope \cite{MAGICsens}.
The blazar remained in a \textit{low state} during most of the monitored period ($\sim$74~h). The average flux F($>$90 GeV) = (2.7$\pm$2.1)$\times$10$^{-11}$ cm$^{-2}$ s$^{-1}$, calculated from the excess at a significance of 4\,$\sigma$, is the lowest VHE \gray\ emission level observed from this source so far. Most notably, the difference between \textit{low state} and \textit{high state} integral photon flux spans an order of magnitude. The measured \textit{high state} fluxes are comparable with the flare detected by MAGIC in October 2017, shortly after the neutrino alert IceCube-170922A.\\

The similarities between the 2018 and 2017 \textit{high states} are also evident in the MAGIC spectral energy distributions shown in Fig. \ref{fig:magic_sed} (right). Both the photon index and flux levels are compatible within errors with the previous measurements. \\

\subsection{Fermi/LAT}
\label{sec:fermi}
In this work, the publicly available $Fermi$/LAT \citep{FermiLAT} data collected in the period from November 1, 2017, to March 1, 2019 (531187205 to 573091205 MET), are used. The Pass 8 SOURCE data were analysed with \text{Fermi} Science-Tools (1.2.1) with {\it P8R3\_SOURCE\_V2} instrument response functions. The 100 MeV - 300 GeV events from a circular region of interest of $12^{\circ}$ around the \gray\ position of \txs\ (RA, Dec)= (77.359,  5.701) were downloaded and analyzed using the standard procedure described by the $Fermi$/LAT Collaboration. The data are binned into pixels of $0.1^{\circ} \times 0.1^{\circ}$ and standard cuts were applied to select the good time intervals ("DATA\_QUAL$>$0" and "LAT\_CONFIG==1", $z_{\rm max}<90^{\circ}$). The model file was created based on the $Fermi$/LAT fourth source catalog \citep[4FGL,][]{4FGL} and it includes all point sources within $17^{\circ}$ from \txs\ and standard templates for Galactic diffuse emission model ({\it gll\_iem\_v07}) and isotropic diffuse emission ({\it iso\_P8R3\_SOURCE\_V2\_v1}).  \\

The \gray\ light curves were calculated by performing an unbinned maximum likelihood analysis with the appropriate quality cuts as described above. 
The photon indexes of all sources, except \txs, as well as the normalization of both background components were fixed to the best-fit values obtained for the whole time period. Only the normalizations of the sources within $12^{\circ}$ from \txs\ were left free during the analysis. Initially the light curve was computed using 7-day intervals and then using the adaptive binning method \citep{2012A&A...544A...6L}. In the latter case, the time bin widths are flexibly adapted to produce bins with constant flux uncertainty above the optimal energies ($E_{\rm 0}$). The adaptively binned light curve computed for 20\% uncertainty and $E_{\rm 0}$ = 216.03 MeV is shown in Fig. \ref{fig:mwl_lc}. The \gray\ flux and photon index variations are investigated using the 7-day binned light curve (Fig. \ref{fig:FermiFluxIndex}).\\

In the low state, the \gray\ spectrum of the \txs\ has been computed by limiting the analysis only to the periods contemporaneous with the MAGIC low state, and the data observed during MJD 58451-58456 (November 29, to December 4, 2018) have been used for the active state. During the analysis, the spectrum of \txs\ has been modeled by a power-law function considering the normalization and index as free parameters. The \gray\ emission spectra in the low and active states are shown in Fig. \ref{fig:SED1}.\\

\subsection{NuSTAR}
\label{sec:nustar}

{\em NuSTAR} \citep{harrison13} observed \txs\  with its two co-aligned X-ray telescopes with corresponding focal planes, focal plane module A (FPMA) and B (FPMB), five times between April 3, 2018, and January 7, 2019, for an exposure time varying between 21.3 ks and 31.7 ks (see Table~\ref{Tab:NuSTAR_0506}). 
The level 1 data products were processed with the {\em NuSTAR} Data Analysis Software (\texttt{nustardas}) package (v2.0.0). Cleaned event files (level 2 data products) were produced and calibrated using standard filtering criteria with the \texttt{NUPIPELINE} task and version 20201101 of the calibration files available in the {\em NuSTAR} CALDB and the OPTIMIZED parameter for the exclusion of the South Atlantic Anomaly passages. Spectra of the sources were extracted from the cleaned event files using a circle of 30\arcsec-radius, while the background was extracted from a nearby circular regions of 70\arcsec-radius on the same chip of the source. The  ancillary  response  files  were  generated  with  the \texttt{numkarf} task,  applying corrections for the point spread function losses, exposure maps and vignetting. The spectra were rebinned with a minimum of 20 counts per energy bin to allow for $\chi^{2}$ spectrum fitting. All errors are given at the 90\% confidence level.\\

The {\em NuSTAR} spectra have been fitted in the 3.0--78 keV energy range with an absorbed power-law with the Galactic absorption corresponding to a hydrogen column density of n$_H$ = 1.11 $\times$ 10$^{21}$ cm$^{-2}$ \citep{2005A&A...440..775K}. The results of the fits are presented in Table~\ref{Tab:NuSTAR_0506}.
The 3-78\,keV flux varied during the {\em NuSTAR} monitoring up to a factor of 2, in particular from (4.54 $\pm$ 0.34) $\times$ 10$^{-12}$ to (7.15 $\pm$ 0.39) $\times$ 10$^{-12}$ erg cm$^{-2}$ s$^{-1}$ between December 8, 2018, and January 7, 2019. No significant variation of the photon index has been observed in the monitored period.\\

\subsection{Swift/XRT}
\label{sec:swiftxrt}
The \textit{Neil Gehrels Swift Observatory (Swift)} has observed the source 41 times between November 8, 2017, and  December 11, 2018. All XRT \citep{SwiftXRT} observations were performed in photon counting mode. The XRT spectra were generated with the {\em Swift} XRT data products generator tool at the UK {\em Swift} Science Data Centre\footnote{http://www.swift.ac.uk/user\_objects} \citep[for details see][]{evans09}. The X-ray spectra in the 0.3--10 keV energy range is fitted by an absorbed power-law model using the photoelectric absorption model \texttt{tbabs} \citep{wilms00} with a HI column density consistent with the Galactic value in the direction of the source, as reported in \citet{2005A&A...440..775K}, i.e. 1.11 $\times$ 10$^{21}$ cm$^{-2}$.
A large number of spectra show low number of counts (i.e. $<$ 200), therefore not allowing us to use the $\chi^{2}$ statistics. To maintain the homogeneity in the analysis, the spectra are grouped using the task \texttt{grppha} to have at least one count per bin and the fit has been performed with the Cash statistics \citep{cash79}. We used the spectral redistribution matrices in the Calibration database maintained by HEASARC. The X-ray spectral analysis was performed using the {\texttt XSPEC 12.9.1} software package \citep{arnaud96}. Single observations with a number of counts $<$ 15, for which it is not possible to constrain the spectral parameters, are not considered. The X-ray spectrum is variable with a photon index, $\Gamma_{\rm\,X}$, varying between 1.27 $\pm$ 0.37 
and 2.71 $\pm$ 0.70. \\

\subsection{Swift/UVOT}
\label{sec:swiftuvot}
The UVOT telescope \citep{UVOT} on board the {\em Swift} satellite observed \txs\ simultaneously with XRT, guaranteeing a coverage in the optical and UV band. Fluxes for each of the six UVOT filters (V, B, U, UW1, UM2, UW2) are calculated with the Heasoft tool \textit{uvotmaghist}, using a 5\arcsec-radius circular region for the source, and a 20\arcsec-radius circular region for the background. The data have been analyzed with the most recent (September 2020) UVOT calibration file.\footnote{see \url{https://www.swift.ac.uk/analysis/uvot/index.php}} The flux measurements are corrected for absorption assuming $E_{B-V}=0.092$ as value for the Galactic extinction \citep{Schlafly:2010dz}. To study the spectral shape in the optical/UV band, and its evolution with time, a power-law fit to the UVOT flux points has been performed for all observations including at least four different filters. All UVOT spectra are characterized by an index softer than $2.0$, indicating that the peak of the synchrotron emission is located below the V band. All power-law indices from UVOT observations are shown in Fig. \ref{fig:XrayFluxIndex}.\\

\subsection{Optical telescopes}
\label{sec:optical}

\txs\ is automatically monitored by the ASAS-SN project \citep{2017PASP..129j4502K}, and the optical light curve has been extracted using the automatic online tool.\footnote{\url{https://asas-sn.osu.edu/}} Figure \ref{fig:mwl_lc} shows the g and V band light curve extracted with an aperture radius of 16\arcsec.\\ 

The KVA measurements were performed contemporaneously with MAGIC for most of the nights, and the data were analysed using the standard procedure, with a signal extraction aperture of 5\arcsec in the R-band \citep{2018A&A...620A.185N}.  \\

The Rapid Eye Mount telescope (REM) observations were triggered after the MAGIC flare. REM is a 60-cm robotic telescope located at the ESO La Silla Observatory \citep{REM1,REM2}. It includes an optical camera with the Sloan filters g, r, i, z and a near-infrared camera equipped with J-H-K filters. Data reduction was carried out following the standard procedures, with the subtraction of an averaged bias frame dividing by the normalised flat frame. The photometric calibration was achieved by using the APASS catalogue.
In order to minimise any systematic effect, we performed differential photometry with respect to a selection of non-saturated reference stars selecting the signal around the centroid of a source within a circle of 10\arcsec-radius.
The monitoring of REM went on for 7 days in r band and the magnitudes observed, converted from the SDSS to Johnson-Cousins photometric system, are comparable with the other measurements in the R-band.\\

The Astronomical Observatory of the University of Siena observed \txs\ in the context of a program devoted to optical photometry of blazars as a follow-up of MAGIC observing campaigns. The instrumentation consists of a remotely operated 30 cm f/5.6 Maksutov-Cassegrain telescope installed on a Comec 10 micron GM2000-QCI German equatorial mount. The detector is a Sbig STL-6303 camera equipped with a 3072 x 2048 pixels KAF-6303E sensor; the filter wheel hosts a set of Johnson-Cousins BVRI filters. Multiple 300s images of \txs\ were acquired in the R band at each visit. After standard dark current subtraction and flat-fielding, images for each visit were averaged and aperture photometry was performed on the average frame by means of the MaximDL software package\footnote{\url{https://diffractionlimited.com/product/maxim-dl/}}, extracting the signal within 7\arcsec\ around the source centroids. The choice of reference and control stars was consistent with the one for the KVA data. The obtained magnitudes have been treated as indicated for the KVA data, as far as correction of galactic extinction, conversion to flux and subtraction of the host galaxy contribution to the measured flux.\\

In order to reproduce the light curve (Fig. \ref{fig:mwl_lc}),  all the optical data are corrected for absorption assuming $E_{B-V}=0.092$.\\

\subsection{Mets\"ahovi}
The 37 GHz observations were performed with the 13.7 m diameter Mets\"ahovi radio telescope.\footnote{\url{https://www.aalto.fi/en/services/metsahovis-main-instruments}} A typical integration time to obtain one flux density data point is between 1200 and 1800 s. The detection limit of our telescope at 37 GHz is on the order of 0.2 Jy under optimal conditions. Data points with a signal-to-noise ratio $<$ 4 are handled as non-detections.

The flux density scale is set by observations of DR~21. Sources NGC~7027, 3C~274 and 3C~84 are used as secondary calibrators. A detailed description of the data reduction and analysis is given in \cite{Terasranta1998}. The error estimate in the flux density includes the contribution from the measurement rms and the uncertainty of the absolute calibration.

\subsection{OVRO}
The 15 GHz data presented here were taken in the frame of the large-scale, fast-cadence monitoring program with the 40 m telescope at the Owens Valley Radio Observatory (OVRO). The data analysis was performed according to a standard analysis described in \cite{2011ApJS..194...29R}.
The flux density of the source increases from about 0.6 Jy in November 2017 to more than twice this value in February 2019. We also note that an increase in the flux was also seen in the 2016 data, well before the neutrino event in September 2017 \citep{Hovatta21}.\\

\begin{figure}[t!]
\centering
\includegraphics[width=\columnwidth]{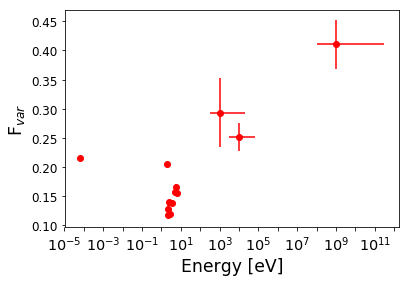}
\caption{Fractional variability parameter for each instrument. All points are calculated using daily averaged fluxes, apart from the point at 1 GeV which uses a 7-days average of $Fermi$/LAT flux.\\}
\label{fig:fvar}
\end{figure}

\section{Variability study} \label{sec:var}

AGNs are known for their variability and \txs\ is no exception. Fig. \ref{fig:mwl_lc} shows that a clear variability is observed in all wavelengths. We use the {\it fractional variability parameter} F$_{var}$ as defined by \cite{2003MNRAS.345.1271V} to quantify the observed trends. The different sensitivity and flux sampling of the different instruments introduce some biases in this method \citep[see e.g.][]{2014A&A...572A.121A}. On the other hand, it is a relatively simple way to quantify and compare the flux variability among the observed energy bands.
Table \ref{tab:fvar} and Fig \ref{fig:fvar} present the F$_{var}$ values for all the instruments contributing to this study, apart from MAGIC. The very low VHE \gray\ activity during these two years yielded low significance flux measurements for most of the single night observations. Therefore the fractional variability in this energy band cannot be estimated. \\

The most reliable variability estimates can be done using the data from monitoring instruments such as $Fermi$/LAT and ASAS-SN. A relatively large data set was also collected with the X-ray and UV instruments on board $Swift$ and the KVA optical telescope. High sensitivity {\em NuSTAR} observations complete the multi-wavelength picture. The most pronounced variability F$_{var}\sim$0.40 is observed in the $Fermi$/LAT \gray\ band, while the X-ray variability is at a lower level of $\sim$0.25-0.30.  The radio, optical and UV bands display a moderate variability of F$_{var}\sim$0.15-0.22. 
A slight spread between the KVA R-band data and the ASAS-SN and UVOT observations at shorter wavelengths is visible. This is probably due to sparser KVA observations, mostly covering high emission states. The radio variability is at a marginally higher level than the optical and UV ones, which is a quite common blazar characteristic.\\

The VHE \gray\ flare observed in December 2018 is very similar to the one observed in October 2017, both in term of the flux level, as well as day-scale variability. In a matter of a few days, we see an order of magnitude flux increment and then a decay.\\

\begin{figure*}[t]
    \centering
    \includegraphics[width=\columnwidth]{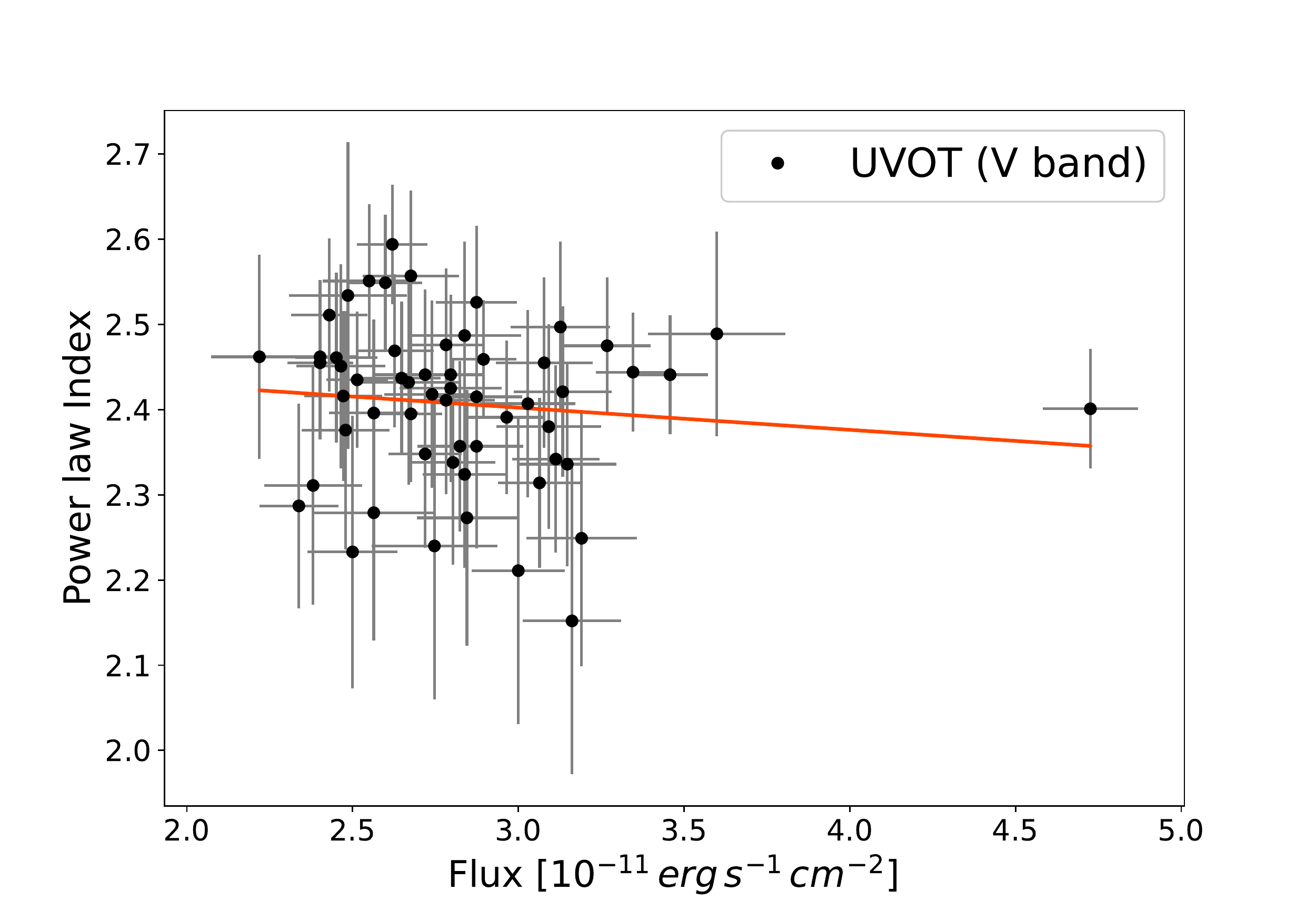}
    \includegraphics[width=\columnwidth]{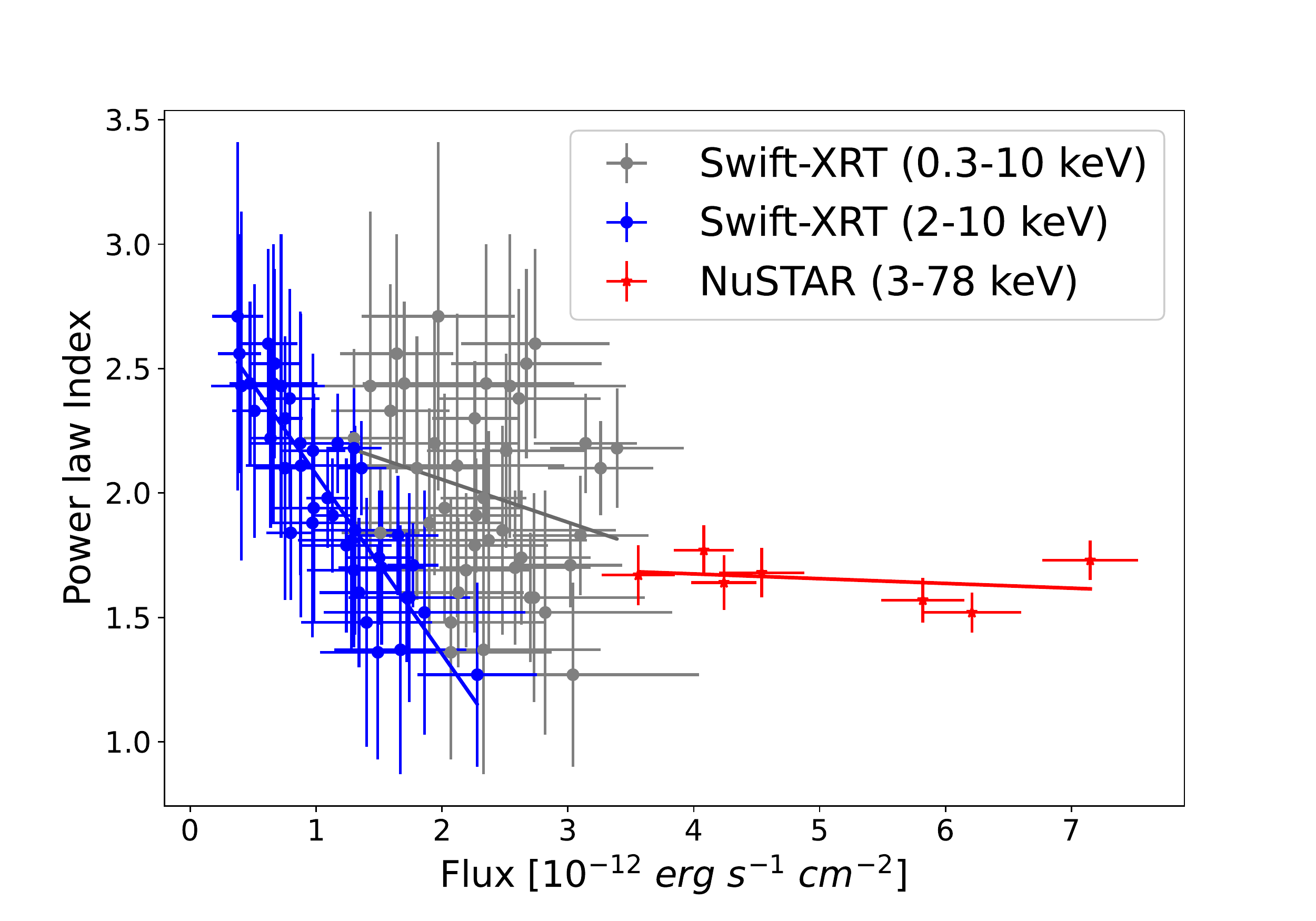}
    \caption {{\it Left}: Correlation between the optical flux in the V band and the photon index in the optical-to-UV band measured by UVOT. {\it Right:} Correlation between the X-ray flux in the 0.3-10, and 2-10\,keV energy bands and the photon indexes measured with {\it Swift}-XRT and that in the 3-78\,keV band as measured with {\it NuSTAR}.  A clear break emerges between the two energetic regimes. In both panels, the dashed lines show the average trend of the data and were obtained with a linear fit.\\ \label{fig:XrayFluxIndex}}
\end{figure*}

\subsection{Spectral variability}
An important aspect to investigate when evaluating blazar variability is the dependence of the synchrotron peak energy or the high-energy peak energy on the flux state. Very often in \gray\ blazars, a harder-when-brighter behaviour, meaning a harder spectrum during enhanced flux states, is observed in both X- and \gray\ bands. This trend is related to the fact that the enhanced flux state can be explained with an injection of high-energy particles, shifting the peaks of the SED towards higher energies and higher flux levels  \citep{10.1051/0004-6361/201935450}. However, some exceptions to this trend are reported in literature  \citep[e.g.,][]{10.1093/mnras/sty2264}.\\

\begin{figure}[t]
    \centering    
     \includegraphics[width=\columnwidth]{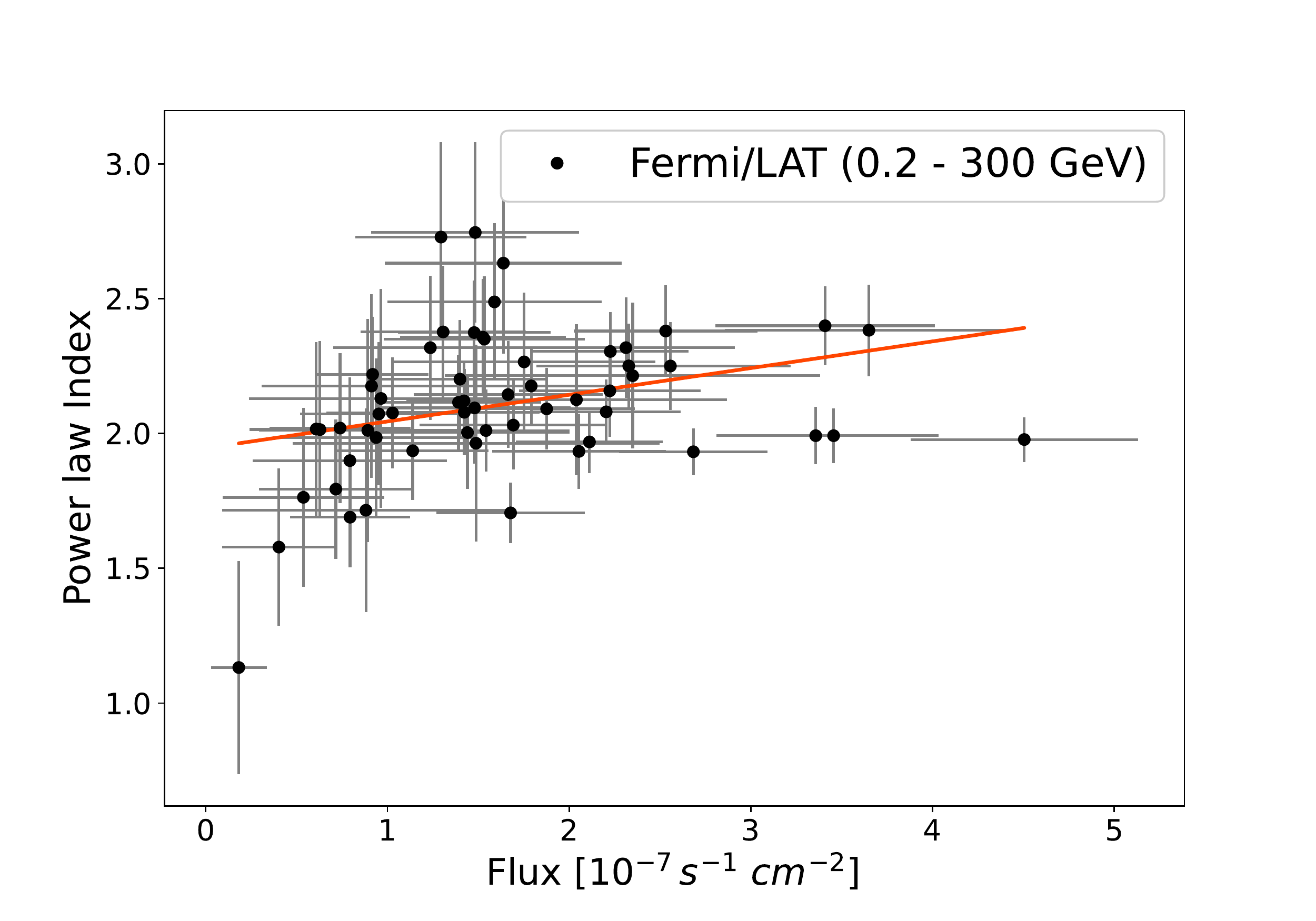}
    \caption{Correlation between the photon flux in the \gray\ band and the photon index measured by {\it Fermi}/LAT.\\      \label{fig:FermiFluxIndex}}
   \end{figure}
   
 To investigate if \txs\ data in different bands follow this trend, a correlation study between the photon flux and the photon index was performed, using the Spearman coefficient $R$. The results of our study are reported in Table~\ref{tab:correlation}. The last column of the table reports the p-value of the hypothesis of no correlation: smaller p-values indicate a higher probability of (anti)correlation. \\

\begin{table}
\caption{Results of the Spearman correlation study between the photon index and the flux level.}
    \centering
    \begin{tabular}{lcc}
\hline
\hline
      Band   & $R$ & p-value \\
     \hline
      UV (V band) & -0.17 & 0.19 \\ 
     X-ray (2--10\,keV)     & -0.90 & $2\times 10^{-15}$ \\
     X-ray (0.3--10\,keV)     & -0.23 & 0.14 \\
     X-ray (3--78\,keV)     & -0.25 & 0.59 \\
     $\gamma$-ray (0.2--300\,GeV) & 0.36 & 0.006 \\ 
     \hline
     \hline
    \end{tabular}
    \label{tab:correlation}
\end{table}

In our analysis, no correlation is found in UVOT data, Figure~\ref{fig:XrayFluxIndex}, left. The corresponding Spearman coefficient is $R=-0.17$, and the p-value is 0.19. In soft X-ray, Figure~\ref{fig:XrayFluxIndex} right panel, a strong anti-correlation between the photon flux in the 2-10\,keV band and the photon index emerges, as supported by a value of $R=-0.90$ (p-value=$2\times 10^{-15}$). This correlation is lost when considering data in the full XRT energy range,  0.3-10\,keV, gray markers in Figure \ref{fig:XrayFluxIndex}, right panel.
No significant spectral variability has been observed in hard X-rays during the {\em NuSTAR} monitoring of the source.
 \\

At higher energies, only {\it Fermi}/LAT data were considered for the study, as the faintness of the signal in the MAGIC energy range prevented a detailed study in this band. The flux-vs-index correlation plot obtained with the {\it Fermi}/LAT data, averaged over 7-days bins, is displayed in Fig. \ref{fig:FermiFluxIndex}. A linear fit to the data suggests a correlation between the two values, indicating a softer-when-brighter trend in $\gamma$ rays. The Spearman coefficient value, however, is only $R=0.36$ (p-value = $0.006)$, so with the available data set we can only conclude that a hint of correlation is deducible in the GeV band.\\

\section{Multi-messenger modeling}
\label{sec:modeling}
\begin{figure*}[t]
\centering
\includegraphics[width=8.5cm]{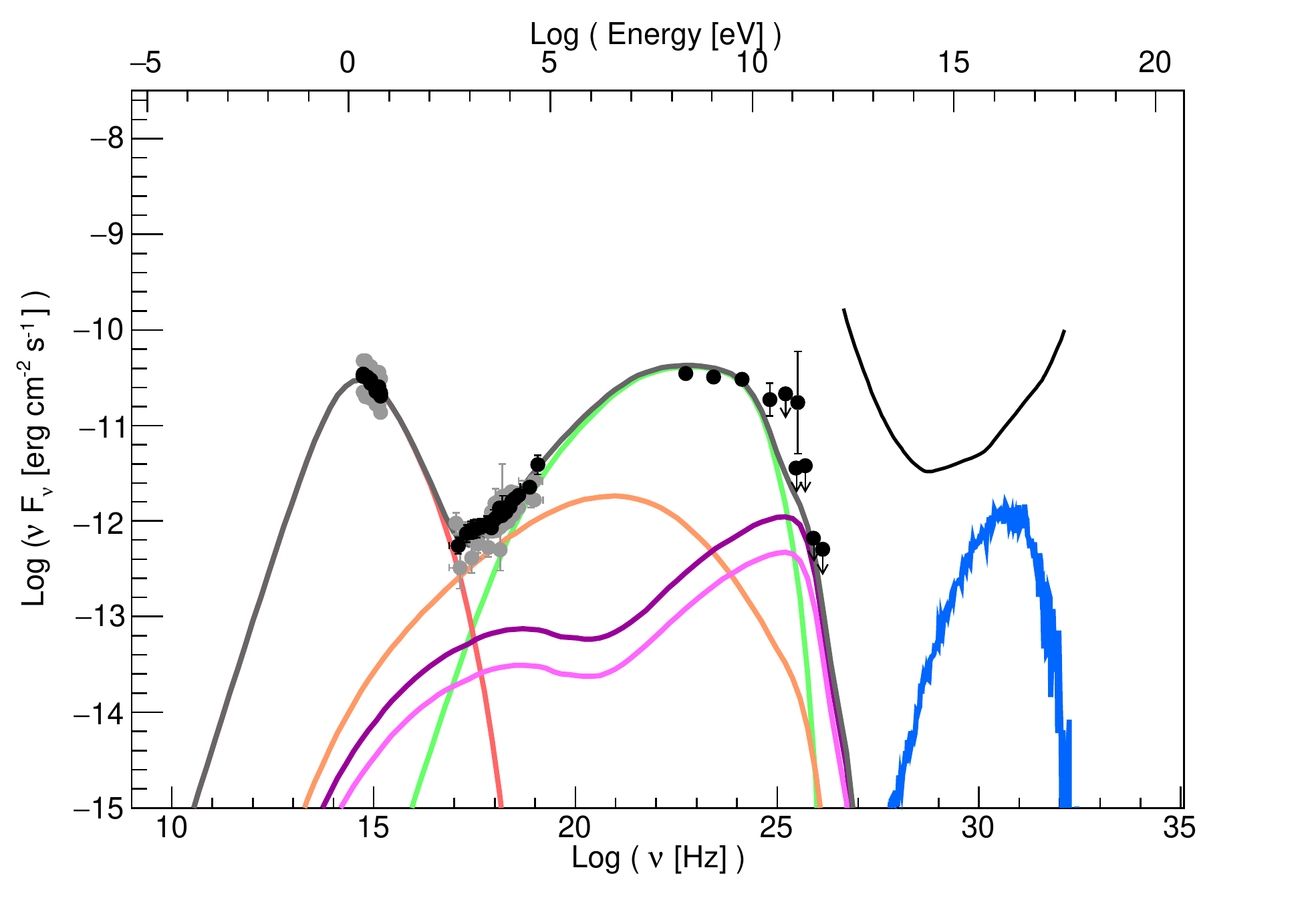}
\includegraphics[width=8.5cm]{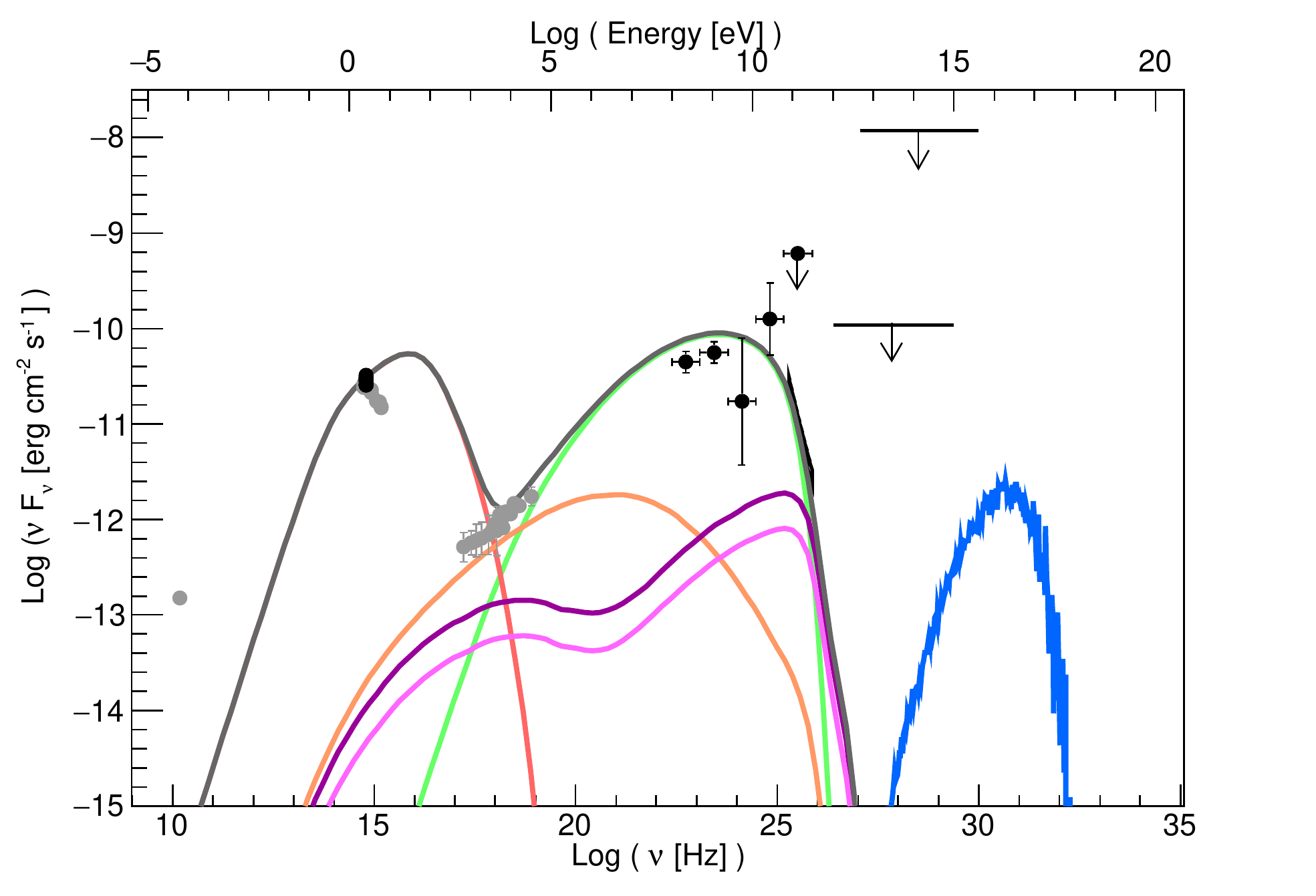}
 \caption{Spectral energy distribution of \txs\ during the 2017-2019 campaign, modeled in a lepto-hadronic scenario. \textit{Left:} low-state (average) during the multi-wavelength campaign. Black data points show the $Fermi$/LAT average spectrum and MAGIC upper limits, and {\it Swift} and {\it NuSTAR} spectra on October 16, 2018. Gray data points show the whole range of optical-to-X-ray spectra during the campaign. \textit{Right:} high-state in VHE \gspaceray s during December 2018. Black data points show the $Fermi$/LAT, MAGIC, and ASAS-SN contemporaneous spectra. Gray data points show the nearest observations in radio, optical, UV and X-rays on December 8, 2018. In both plots, the solid lines represent synchrotron emission by primary electrons (red); inverse-Compton emission over photons from the jet layer (green); Bethe-Heitler cascade (orange); cascade from $\pi^0$ (violet) and $\pi^\pm$ (pink) decay; neutrino emission (blue). In the left plot we show the IceCube $5\sigma$ sensitivity curve for point-source searches \citep[for eight years exposure and $0^\circ$ declination, see][]{IC8years}. In the right plot we show the IceCube and ANTARES flux upper limits (at lower and higher energies, respectively) following the detection of a VHE flare in December 2018.\\}
 \label{fig:SED1} 
\end{figure*}

The SEDs of \txs\ during the 2017-2019 multi-wavelength campaign are shown in Fig. \ref{fig:SED1}, separately for the VHE flare of December, 2018, and the average low-state. For the latter, we use the \textit{low-state} spectra from {\em Fermi}/LAT and MAGIC, and {\em NuSTAR} and {\em Swift} data from October 16, 2018 as an exemplary spectrum for the low-state (together with, in gray, the whole range of optical-to-X-ray observations). For the VHE high-state, the only instrument performing contemporaneous observations with MAGIC were $Fermi$/LAT and ASAS-SN. We include nonetheless the nearest multi-wavelength observations in radio, optical, UV and X-rays (on December 8, 2018). Blazar SEDs can be fit by both leptonic and hadronic models. Neutrino detection from a blazar can indeed be the key to discriminate between the two emission scenarios. Fully aware that in absence of neutrinos the SED of \txs\ can be solely described by leptons, we model the SEDs in a lepto-hadronic scenario inspired by the September 2017 photon-neutrino event: the dominant radiative process in the \gray\ band is external-inverse-Compton, with hadronic radiative components (Bethe-Heitler and pion-decay pair-cascade) emerging in the X-ray and VHE bands. The dominant low-energy photon field for proton-photon interactions is synchrotron radiation from the jet layer that encompasses the jet spine in which the plasmoid moves \citep{Ghisellini05}. This geometry is also supported by recent spatially resolved radio observations described in \citet{Ros20}.\\

The numerical simulation is performed with the code described in \citet{Cerruti15}, expanded to allow for hadronic interactions over arbitrary external photon fields. Photo-meson interactions in the code are implemented via the Monte-Carlo code SOPHIA \citep{Sophia}, while Bethe-Heitler pair-production is implemented following \citet{Kelner08}. Absorption of VHE \gspaceray s on the extragalactic background light is implemented using the model by \citet{Franceschini08}. The pair-cascades associated with hadronic interactions are computed iteratively generation-by-generation. This assumes that the cascades are not self-supported, and that the dominant photon fields are either the external target photon field, or the synchrotron emission by primary electrons, but never the photon emission from the cascades themselves. The radiative code computes the steady-state emission from a spherical plasmoid in the jet, filled by a tangled, homogeneous magnetic field, and interacting with external photons from the jet layer. The plasmoid is fully characterized by three free parameters: the bulk Lorentz factor $\Gamma$ (translated into the Doppler factor $\delta$ assuming an angle to the line of sight $\theta_{\rm view}$), the radius $R^\prime$, and the magnetic flux density $B^\prime$.\footnote{Here and in the following, primed quantities are given in the reference frame of the plasmoid. Double primed quantities are in the reference frame of the layer.} The emitting region is filled with a population of primary electrons and protons, without any assumption on the underlying acceleration mechanism. The primary particle distributions are parametrized by broken-power-law functions with exponential cut-offs at their maximum Lorentz factor $\gamma^\prime_{\rm{Max}}$, and are thus characterized by six free parameters each (the three Lorentz factors $\gamma^\prime_{\rm{min}}$, $\gamma^\prime_{\rm{break}}$, $\gamma^\prime_{\rm{Max}}$, the two indexes $\alpha_1$, and $\alpha_2$, and the normalization $K^\prime$), for a total of nine free parameters.
In addition, the external photon field from the jet layer carries other eight free parameters: the layer region is a hollow cylinder (parametrized by its inner and outer radii $R^{\prime\prime}_{\rm in,layer}$ and $R^{\prime\prime}_{\rm out,layer}$, and its height $H^{\prime\prime}_{\rm layer}$), moving with bulk Lorentz factor $\Gamma_{\rm layer}$ and filled with a homogeneous magnetic field $B^{\prime\prime}_{\rm layer}$, and an electron population parametrized by a broken power-law function with its six free parameters. Such a model is clearly degenerate, and several assumptions have to be introduced to reduce the number of free parameters: the primary electrons and protons are considered to be co-accelerated, sharing the same index $\alpha_{1,\rm{e}} = \alpha_{1,\rm{p}}$; the primary protons are not cooled (i.e. all cooling processes become relevant at Lorentz factors higher than $\gamma^\prime_{\rm{p,Max}}$) and thus the proton distribution is parametrized by a simple power-law function; the minimum proton Lorentz factor is fixed to $1$, to avoid biasing the total luminosity estimate; the Doppler factor of the plasmoid is fixed to a rather typical value of $40$ \citep{Tavecchio10, Zhang12}, with an angle to the line of sight $\theta_{\rm view} = 0.8^\circ$; the geometry of the layer is $R^{\prime\prime}_{\rm in,layer} = R^\prime$, $R^{\prime\prime}_{\rm out,layer} = 1.5 R^\prime$, while the height of the layer in the frame of the spine is $H^\prime_{\rm layer} = H^{\prime\prime}_{\rm layer} / (\Gamma \Gamma_{\rm layer} (1 - \beta \beta_{\rm layer})) = R^\prime$ \citep[see][]{Ghisellini05}; the photon field from the layer is computed using $\delta_{\rm layer}  = 4$, $B^\prime_{\rm layer} = 1.8\ \textrm{G}$, $\alpha_{\rm e,1,layer} = 2.0$, $\alpha_{\rm e,2,layer} = 2.9$, $\gamma^\prime_{\rm e,min,layer} = 290$, $\gamma^\prime_{\rm e,break,layer} = 2000$,  and $\gamma^\prime_{\rm e,Max,layer} = 3.9\times10^5$ \citep[identical to the values used in][]{MAGIC_TXS_paper2}. The normalization of the synchrotron emission from the layer is left free to vary and is expressed as photon energy density in the reference frame of the plasmoid, $u^\prime_{\rm ph,layer}$. The external inverse-Compton component is corrected by a factor $\delta/\delta_{\rm layer}$ following \citet{Dermer95}. These physical constraints reduce the number of free parameters of the model to eleven: $R^\prime$, $B^\prime$, $\gamma^\prime_{\rm{e,min}}$, $\gamma^\prime_{\rm{e,break}}$, $\gamma^\prime_{\rm{e,Max}}$, $\gamma^\prime_{\rm{p,Max}}$, $\alpha_1$, $\alpha_{2\rm{,e}}$, $K_{\rm{e}}$, $K_{\rm{p}}$, and $u^\prime_{\rm ph,layer}$. This number remains larger than the number of independent observables, so no numerical fitting of the model to the data is performed. We limit ourselves to identify a lepto-hadronic solution (among many) that can reproduce the observations.\\

\begin{deluxetable}{lcc}
\tablecaption{Parameters of the lepto-hadronic models \label{table:model}}
\tablewidth{0pt}
\tablehead{
\colhead{} & \colhead{Low-state} & \colhead{High-state} 
}
\startdata
        $z$ & $0.337$ & $''$\\
	    $\theta_{\rm view}$ [deg]  & $0.8$ & $''$\\
 		$\delta$ & $40$ & $''$ \\
 	    $^\star \Gamma $ & $22$ & $''$\\
 $R^\prime$ [10$^{16}$ cm] & $1.1$ & $''$ \\
 $^\star \tau_{\rm var}$ [hours] & $3.4$ & $''$\\
 		\hline
 		 $B^\prime$ [G] & $1$ & $''$ \\
 		$^\star u^\prime_B$ [erg cm$^{-3}$] & $0.04$ & $''$ \\
 		\hline
 		$\gamma_{\rm e,min}^\prime $& $800$ &  $''$  \\
 		$\gamma_{\rm e,break}^\prime $& $2.2\times10^3$ & $1\times10^4$   \\
 		$\gamma_{\rm e,max}^\prime$& $10^4$ & $2\times10^4$\\
 		$\alpha_{\rm e,1}=\alpha_{p,1}$ & $2.0$ & $''$ \\
 		$\alpha_{\rm e,2}$ & $4.0$ & $''$  \\
 		$K^\prime_{\rm e}$ [cm$^{-3}$] & $10^3$ & $520$\\
 		$^\star u^\prime_{\rm e}$ [erg$\,$cm$^{-3}$] & $1.0\times10^{-3}$ & $1.0\times10^{-3}$\\
 		\hline
 		$\gamma_{\rm p,min}^\prime$& $1$ & $''$ \\
 		$\gamma_{\rm p,max}^\prime [10^7]$& $1$ & $''$\\
        $K_{\rm p}^\prime$ [cm$^{-3}$]  & $3.0\times10^{3}$ & $5.0\times10^{3}$ \\
 		$^\star u^\prime_{\rm p}$ [erg$\,$cm$^{-3}$] &$70$ &  $120$ \\
 		\hline
 		$\delta_{{\rm layer}}$ & $4$ & $''$ \\
 		$^\star \Gamma_{{\rm layer}}$ & $2.2$ & $''$ \\
 		$R^\prime_{{\rm in,layer}}$ [10$^{16}$ cm] & $1.1$ & $''$ \\
 		$B_{\rm layer}^{\prime\prime}$ [G] & $1.8$ & $''$ \\
 		$\gamma_{\rm e,min,layer}^{\prime\prime}$& $290$ & $''$  \\
 		$\gamma_{\rm e,break,layer}^{\prime\prime}$& $2000$ & $''$  \\
 		$\gamma_{\rm e,max,layer}^{\prime\prime}$& $3.9\times10^5$ & $''$\\
 		$\alpha_{\rm e,1,layer}$ & $2.0$ &  $''$ \\
 		$\alpha_{\rm e,2,layer}$ & $2.9$ &  $''$ \\
 		$u^\prime_{\rm ph,layer}$ [erg cm$^{-3}$] &$1.5\times10^{-2}$ & $''$  \\
 		\hline
 		$^\star u^\prime_{\rm e}/u^\prime_{\rm B} [10^{3}]$ & $0.025$ & $0.024$\\
    	$^\star u^\prime_{\rm p}/u^\prime_{\rm B} [10^{3}]$ & $1.7\times10^3$ & $2.9\times10^3$\\
 		$^\star L$ [10$^{47}$ erg s$^{-1}$] & $7.6$ & $12.7$  \\
 		$^\star \nu$ [yr$^{-1}$] & $0.14$ &  - \\
 	    $^\star \nu_{\rm{GOLD}}$ [day$^{-1}$] & - & $2.1\times10^{-4}$  \\
 		\hline
 		\hline
 		\enddata
 	 \tablecomments{ The quantities flagged with a star ($^\star$) are derived quantities, and not model parameters. The luminosity of the emitting region has been calculated as \mbox{$L=2 \pi R^{\prime 2}c\Gamma^2(u^\prime_{\rm B}+u^\prime_{\rm e}+u^\prime_{\rm p})$}, where $u^\prime_{\rm B}$, $u^\prime_{\rm e}$, and $u^\prime_{\rm p}$ are the energy densities of the magnetic field, the electrons, and the protons, respectively. For the high-state model, only the parameters that change with respect to the low-state one are indicated.}
 		\end{deluxetable}

The multi-wavelength observations and the variability studies presented in the previous sections provide further constraints to the modeling. The position of the synchrotron peak is located in the optical band or lower, as determined by UVOT. The X-ray band marks the transition between two different emission components, as clearly shown by the different spectral variability properties in {\em Swift}/XRT and {\em NuSTAR} data. In the lepto-hadronic scenario, this is naturally explained by the transition from the synchrotron component to the inverse-Compton one, with additional contribution from the Bethe-Heitler pair cascade. We start by studying the low-state emission of \txs. The free parameters of the model are first optimized to get a leptonic modeling of the optical-UV and \gray\ part of the SED. The proton population is then increased until the emission from Bethe-Heitler and photo-meson cascades starts violating X-ray and VHE observations. An additional constraint comes from IceCube observations, that identify a hotspot at the position of \txs\, corresponding to 12.3 events in ten years. We compute the expected neutrino rate by convolving the neutrino spectrum from the model with the IceCube effective area for point-source searches. The lepto-hadronic solution, shown in Figure \ref{fig:SED1} (with model parameters provided in Table \ref{table:model}), has a hadronic contribution emerging in the X-rays and at VHE as Bethe-Heitler and pion-decay cascade. The model for the low state predicts an IceCube neutrino rate of 1.4 events in ten years (estimated using the IceCube point-source effective area), much less than the recent results by \citet[][12.3 neutrinos in ten years]{Aartsen2010years}. This supports the idea that the hotspot at the position of \txs\ in IceCube data is dominated by flaring events (i.e. the flare which occurred in 2014-15 and the high-energy neutrino of September 2017). \\

It is important to underline that the hadronic model remains degenerate unless explicit assumptions on the acceleration/cooling processes (and thus the maximum proton Lorentz factor $\gamma_{\rm p,Max}^\prime$) are made. The maximum proton energy is related to the peak energy of the neutrino emission: lower values of $\gamma_{\rm p,Max}^\prime$ would imply a higher neutrino rate (because the spectrum would then peak closer to the maximum of IceCube sensitivity) at the expense of a higher total jet power. In an opposite way, the total jet power (in our model equal to $7.6\times10^{47}$ erg s$^{-1}$, or $7.6-76$ $L_{Edd}$ for a super-massive black-hole with mass $M_\bullet = 10^{9-8} M_\odot$ ) can be reduced if the maximum proton energy is increased, resulting in a shift towards higher energies of the neutrino peak. The total jet power is very dependent on the index of the proton distribution and on $\gamma_{\rm p,min}^\prime$. The model presented here, with $\gamma_{\rm p,min}^\prime = 1$ and $\alpha_{\rm{p}} = 2.0$ is explicitly conservative in this sense: a harder particle index, or $\gamma_{\rm p,min}^\prime \gg 1$ will significantly lower the total jet power.\\

Once we obtained a satisfactory modeling of the low-state emission of \txs, we investigated the December 2018 VHE flare. In this case the only truly simultaneous observations are from \magic, \lat, and ASAS-SN: the behaviour of the synchrotron component of the SED during the flare is unknown.
On December 3, 2018 the MAGIC collaboration issued an alert \citep{MagicATEL2018} to encourage further multi-wavelength observations of \txs. Together with the X-ray and optical instruments, both IceCube and ANTARES set upper limits on the neutrino emission in the week/10 days around the VHE \gray\ flare \citep{AntaresATel, ICAtel}. They are also shown on the SED plot. The model shown in Figure \ref{fig:SED1} (with model parameters in Table \ref{table:model}) is obtained through a minimal variation of the better constrained low-state model: the only parameters that change are the ones from the primary particle populations: the electron break Lorentz factor is increased to $10^4$, while the maximum Lorentz factor is increased to $2\times10^4$ (while maintaining the same electron energy density); the primary proton distribution is unchanged in spectral shape, but increased in overall normalization by about a factor of two. The model for the \gray\ flaring state predicts the presence of an optical/UV/soft-X-ray flare that was missed due to absence of simultaneous observations in that band. We highlight that this statement is very model-dependent, and there exist other solutions (obtained assuming that other model parameters are driving the variability) with a very different behaviour in the synchrotron and neutrino components. In this scenario the expected neutrino rate in IceCube data \citep[computed using the GOLD event effective area,][]{2019ICRC...36.1021B} is $2.1\times10^{-4}$ events in a day, or $1.5\times10^{-3}$ in a week, completely consistent with the non-detection of neutrinos associated with this \gray\ flare.\\

\section{Summary and conclusions}
\label{sec:concl}
 We have performed an extensive multi-wavelength campaign from radio to VHE \gspaceray s on the {\it "neutrino blazar"} candidate TXS 0506+056. The observations cover the time span from November 2017 (right after the neutrino event of September 2017) to February 2019. The source has been very poorly studied before 2017, and this work represents the most complete characterization of the behaviour of \txs\ on a time-scale of a few years.\\

In the VHE band, the source showed another day-long flare on December 1-3, 2018, observationally very similar (both in flux level and photon index) to the one of September 2017. In the rest of the multi-wavelength campaign, \txs\ was not detected neither on single days, nor in short periods of time. When aggregating all data collected with MAGIC, which amounts to 75 hours, one gets a VHE \gray\ excess at the level of 4\,$\sigma$.  No neutrino flares associated with the VHE flare in December 2018 have been reported. In the HE \gray\ band, \txs\ was in a very different state compared to that from September 2017: while at that time the source was in a long-term (month-scale) brightening, during this campaign the source showed a much lower average state with much faster (days-to-week) flaring activity. No significant flares are detected in X-rays or optical/UV. In the radio band, OVRO observations at 15 GHz revealed the presence of a long-term brightening of the source which increased its flux by around a factor of two throughout the campaign. The same brightening is seen by Mets\"ahovi at 37 GHz, with the same approximate two-fold increase in flux during the discussed time period. While the radio behaviour seems uncorrelated with the other bands on the time-scale of the observing campaign, it is important to underline that the physical zones sampled by radio observations are different from the ones at higher energies, and increases in radio flux densities have been associated to \gray\ flaring activity \citep{2001ApJ...556..738J,Lahteen03,2011A&A...532A.146L}, and, tentatively, with neutrinos \citep{Hovatta21}.
Spectral studies reveal only marginal spectral variability in UVOT and $Fermi$/LAT, indicating that the position of the SED peaks was stable during the campaign. On the other hand, a very interesting spectral variability is seen in X-rays: while the hard-X-ray band (3-70 keV, sampled by {\em NuSTAR}) shows no spectral variability, the soft-X-rays are much more variable, with a clear harder-when-brighter behaviour detected in the 2-10 keV band with XRT observations. When studying the whole XRT band (0.3-10 keV), the anti-correlation between the photon index and the flux is lost, suggesting the presence of a transition between different radiation mechanisms in the keV band, that in a lepto-hadronic model corresponds to the transition from synchrotron radiation by primary electrons, through the one by Bethe-Heitler pairs, to inverse-Compton. \\

We perform a lepto-hadronic modeling of \txs, starting from the low-state of the source which is now, for the first time, very well characterized. A scenario in which the emission is dominated by the leptonic component (inverse-Compton in the \gray\ band), with hadronic interactions happening between protons in the blazar region and photons from the jet-layer, can provide a good description of the SED. The model predicts a detection rate by IceCube of 0.14 neutrinos per year, or 1.4 neutrinos in ten years of observations, consistent with the most recent estimates by IceCube (12.3 neutrinos in ten years). This modeling strengthens the idea that the neutrino emission from AGNs is dominated by their flaring events. The VHE flare of December 2018 does not have simultaneous observations at lower energies, which prevents us to efficiently constrain the theoretical model. We present a tentative modeling of these observations assuming that the only parameters that changed with respect to the low-state of the source were the electron and proton primary distributions. The conclusion is that the \gray\ flare had to be accompanied by a brightening in the optical/UV/X-ray band. Within this model, the associated neutrino emission would have been too rapid and not bright enough to be detected by IceCube (predicted rate of $2.1\times10^{-4}$ neutrinos per day).\\

The association of the neutrino IceCube-170922A with the flaring blazar \txs, has promoted this once {\it vanilla blazar} to one of the most important \gray\ sources in the sky. To fully understand the multi-messenger gamma-neutrino association, a thorough characterization of the source is needed. In this work we presented the largest multi-wavelength campaign on \txs\ to date, with 16 months of data-taking from radio to VHE, but future observations are definitely needed. It is of particular importance the characterization of the variability properties of the source and its duty cycle in the various energy bands, to ultimately quantify how exceptional the 2017 \gray\ flare was. Several questions remain open, and new one have arisen from this campaign. While the source's behaviour in the GeV band changed significantly from September 2017, we detected a new VHE flare very similar to the previous one. Another highlight of this campaign is the radio behaviour of \txs, which is undergoing a long-term brightening completely uncorrelated to any other wavelength. Future multi-wavelength and multi-messenger observations of \txs\ will help cast light on hadronic acceleration in AGNs jets, and on the role of AGNs as cosmic rays and neutrino sources.\\

\acknowledgments
We would like to thank the Instituto de Astrof\'{\i}sica de Canarias for the excellent working conditions at the Observatorio del Roque de los Muchachos in La Palma. The financial support of the German BMBF, MPG and HGF; the Italian INFN and INAF; the Swiss National Fund SNF; the ERDF under the Spanish Ministerio de Ciencia e Innovaci\'on (MICINN) (PID2019-104114RB-C31, PID2019-104114RB-C32, PID2019-104114RB-C33, PID2019-105510GB-C31,PID2019-107847RB-C41, PID2019-107847RB-C42, PID2019-107847RB-C44, PID2019-107988GB-C22); the Indian Department of Atomic Energy; the Japanese ICRR, the University of Tokyo, JSPS, and MEXT; the Bulgarian Ministry of Education and Science, National RI Roadmap Project DO1-400/18.12.2020 and the Academy of Finland grant nr. 320045 is gratefully acknowledged. This work was also supported by the Spanish Centro de Excelencia ``Severo Ochoa'' (SEV-2016-0588, SEV-2017-0709, CEX2019-000920-S), the Unidad de Excelencia ``Mar\'{\i}a de Maeztu'' (CEX2019-000918-M, MDM-2015-0509-18-2) and by the CERCA program of the Generalitat de Catalunya; by the Croatian Science Foundation (HrZZ) Project IP-2016-06-9782 and the University of Rijeka Project 13.12.1.3.02; by the DFG Collaborative Research Centers SFB823/C4 and SFB876/C3; the Polish National Research Centre grant UMO-2016/22/M/ST9/00382; and by the Brazilian MCTIC, CNPq and FAPERJ.
E.P. acknowledges funding from Italian Ministry of Education, University and Research (MIUR) through the "Dipartimenti di eccellenza” project Science of the Universe. M.C. has received financial support through the Postdoctoral Junior Leader Fellowship Programme from la Caixa Banking Foundation, grant No. LCF/BQ/LI18/11630012.
This research has made use of data from the OVRO 40-m monitoring program which was supported in part by NASA grants NNX08AW31G, NNX11A043G and NNX14AQ89G, and NSF grants AST-0808050 and AST-1109911, and private funding from Caltech and the MPIfR.
This publication makes use of data obtained at Mets\"ahovi Radio Observatory, operated by Aalto University in Finland.\\

\facilities{MAGIC, $Fermi$/LAT, NuSTAR, Swift, KVA, REM, ASAS-SN, Mets\"ahovi, OVRO}

\newpage

\appendix
\section{Multiwavelength data tables}

\begin{table}[ht!]
\caption{Log and fitting results of {\em NuSTAR} observations of \txs\ using a power-law model model with $N_{\rm H}$ fixed to Galactic absorption,  i.e. 1.11$\times$10$^{21}$ cm$^{-2}$.} 
\label{Tab:NuSTAR_0506}
\begin{center}
\begin{tabular}{ccccc}
\hline
\hline
\multicolumn{1}{c}{\textbf{Date}} &
\multicolumn{1}{c}{\textbf{MJD}} &
\multicolumn{1}{c}{\textbf{Net exposure time}} &
\multicolumn{1}{c}{\textbf{Flux$_{\rm\,3.0-78\,keV}$}} &
\multicolumn{1}{c}{\textbf{Photon index}}  \\
\multicolumn{1}{c}{[UT]} &
\multicolumn{1}{c}{} &
\multicolumn{1}{c}{[s]} &
\multicolumn{1}{c}{[10$^{-12}$ erg cm$^{-2}$ s$^{-1}$]} &
\multicolumn{1}{c}{[$\Gamma_{\rm\,X}$]}  \\
\hline
2018-04-03 &  58211 & 26030  &  5.82 $^{+0.33}_{-0.39}$ & 1.57 $\pm$ 0.09 \\
2018-10-16 &  58407 & 31868  &  6.21 $^{+0.39}_{-0.35}$ & 1.52 $\pm$ 0.08 \\
2018-11-15 &  58437 & 26880  &  4.24 $^{+0.26}_{-0.28}$ & 1.64 $\pm$ 0.11 \\
2018-12-08 &  58460 & 25899  &  4.54 $^{+0.34}_{-0.34}$ & 1.68 $\pm$ 0.10 \\
2019-01-07 &  58490 & 25889  &  7.15 $^{+0.38}_{-0.40}$ & 1.73 $\pm$ 0.08 \\
\hline
\hline
\end{tabular}                
\end{center}
\end{table}

\begin{table}[ht!]
\small
\caption{Fractional variability of \txs}

\label{tab:fvar}
\centering
\begin{tabular}{c|ccccccccc}
\hline
\hline      
\textbf{Instrument} & \textbf{Fermi/LAT} & \textbf{NuSTAR} & \textbf{Swift/XRT} & \textbf{Swift/UVOT} & \textbf{KVA} & \textbf{ASAS-SN} & \textbf{OVRO}\\
 \hline
F$_{var}$ &  0.410 $\pm$ 0.045 & 0.25$\pm$0.02 & 0.30$\pm$0.06 &0.128$\pm$0.007 (V) & 0.194$\pm$0.002 & 0.143$\pm$0.001 (g) &  0.216 $\pm$ 0.002 \\
  & & & & 0.120$\pm$0.005 (B) & & 0.136$\pm$0.004 (V) & \\
  & & & &0.138$\pm$0.005 (U) & & &\\
   & & & &0.157$\pm$0.005 (W1) & & &\\
    & & & & 0.166$\pm$0.005 (M2) & & &\\
    & & & & 0.156$\pm$0.004 (W2) & & &\\
    \hline
    \hline
 \end{tabular}
\end{table}

\bibliography{references}{}
\bibliographystyle{aasjournal}

\end{document}